\documentclass[11pt]{article}

\usepackage{xcolor}
\usepackage{microtype}
\usepackage{graphicx}
\usepackage{lineno}
\usepackage{hyperref}
\usepackage{authblk}
\usepackage{geometry}

\begin{document}

\title{Safe liquid scintillators for large scale detectors}

\author{A.~Bonhomme, C.~Buck, B.~Gramlich, M.~Raab}
\affil{Max-Planck-Institut f\"ur Kernphysik, Saupfercheckweg 1, 69117 Heidelberg, Germany}




\maketitle
\flushbottom

\abstract{Many experiments in particle physics, in particular in the field of neutrino searches, rely on organic liquid scintillators as target and detection material. The size of these detectors was continously growing in the last decades, up to the kiloton scale. In several cases these detectors are located at sites with enhanced safety requirements as underground laboratories or in the vicinity of nuclear reactors. Therefore, there is strong demand in liquids which are safe with respect to aspects as fire protection, human health or environmental pollution. In particular, properties as the flash point, the vapor pressure or the toxicity need to be significantly improved as compared to classical solvents such as xylene or pseudocumene. We present and compare the performance and optical properties of scintillators based on high flash point solvents. In particular polysiloxane based scintillators are characterized by outstanding properties in terms of safety.}

\section{Introduction}
\label{sec:intro}

Organic liquid scintillators (LS) have been used in neutrino detectors early on, starting with the pioneering neutrino experiment of Reines and Cowan~\cite{Reines_01} close to a nuclear reactor. Many other neutrino experiments followed, some of them involving metal loading. The main advantages of such LS detectors as compared to other technologies are the rather low energy threshold, detector homogeneity, availability of several purification techniques, ﬂexible handling, scalability to large volumes and cost-effectiveness. For example, LS experiments such as Borexino~\cite{Borex} or KamLAND~\cite{Kamland} reached unprecedented radiopurity levels for liquid amounts in the kiloton scale. With the help of Gd-loaded LS the reactor neutrino experiments Double Chooz~\cite{DC}, Daya Bay~\cite{DB} and RENO~\cite{RENO} were able to determine the neutrino mixing angle $\theta_{13}$. In the future we also expect new groundbreaking results from projects as JUNO~\cite{JUNO} or SNO+~\cite{SNO}, which even surpass the volumes of the aforementioned experiments.

However, there is also a big challenge whenever LS are involved which is related to the safety aspects. First, many of the candidate solvents for such liquids are highly flammable. Second, most of these organic materials are hazardous for the human health or even toxic. Last but not least there is a risk of liquid spills, in particular during transport, purification or filling of the detectors. Since some of the involved chemicals are environmental hazards, careful precautions have to be taken to prevent any liquid loss, in particular into ground water or the natural environment. In recent years the focus shifted to solvents with high flash points, often named as "safe scintillators". Although there is nowadays a good set of candidates with significantly reduced fire risk, many of the solvents are still classified as hazardous to human health or the environment. Another issue is often the characteristic smell and the non-negligible vapor pressure leading to severe safety concerns. These are enhanced when operations are performed with open vessels or at specific sites. Examples are nuclear power plants, which are strong neutrino sources, or underground laboratories, which are often required to reduce background events from cosmic radiation such as muons.          

In this article, we compare and characterize LS based on high flash point solvents. The advantages and disadvantages are discussed. In chapter 2, we describe the scintillator composition and list appropriate materials for the LS production along with some basic properties. The experimental setup and the measurement procedures are explained in chapter 3. The results of the performed measurements on the light yield, the capabilities of background discrimination using pulse shape information and transparency are summarized in chapter 4.    

\section{Scintillator composition and materials}
\label{sec:production}

An organic LS typically consists of at least two components. The basis is an aromatic solvent. The fluorescent solvent molecules are partly excited by ionizing particles passing through the liquids. These excitations mainly involve the delocalized electrons of the $Pi$-bonds in the phenyl groups. For an isolated fluorescent solvent molecule, the deexcitation to the ground state, which is typically a singlet state, would be accompanied by the emission of a photon on a time scale of about ten nanoseconds. In a pure aromatic solvent most of these ﬂoursecent molecules would be immediately absorbed. Since the quantum yield for most solvent molecules is less than 50\%~\cite{MLS}, the pure solvent would not emit a significant amount of scintillation light.

Therefore, an additional ﬂuorescence molecule has to be added, the primary wavelength shifter (ﬂuor). Its purpose is to shift the emission spectrum towards longer wavelengths where the liquid is more transparent. Energy is mostly transfered radiationless from the solvent to the ﬂuor. Depending on the wavelength region of maximal sensitivity of the photo sensors or in case of large scale detectors there might be the need for an additional secondary wavelength shifter. Here, the scintillation light is even further shifted to maximize detection efficiency and minimize self-absorption effects.  

\subsection{Solvent candidates}

The simplest form of a scintillator solvent candidate is benzene. Benzene has some of the common features which are found in many aromatic solvents used for LS: clear and colorless liquid, characteristic odour, density slightly below 1\,kg/l, refractive index close to 1.5, high vapor pressure, highly flammable and hazardous. Benzene derivatives with very similar properties and favorable scintillation characteristics are e.g.~toluene, p-xylene, pseudocumene (PC) or anisole. Since all of them are classified as flammable and due to their rather high vapor pressure, alternative safer solvents are nowadays considered for most running and upcoming experiments. Some of them are described below and listed along with their main properties in table~\ref{tab1}. 

\begin{table}[h]
\caption[Solvents]{Density, flash point, vapor pressure p, refractive index n and viscosity are shown for different solvent candidates. \label{tab1}}
\begin{center}
\begin{tabular}{lcccccc}
Molecule & formula & density [kg/l] & flash point & p [hPa] & n & visc.~[mm$^2$/s] \\
\hline
Toluene & C$_7$H$_{8}$ & 0.87 & 4$^\circ$C & 29 & 1.50 & 0.7 (20$^\circ$C) \\
Xylene & C$_8$H$_{10}$ & 0.86 & 27$^\circ$C & 9 & 1.50 & 0.7 (25$^\circ$C) \\
PC & C$_9$H$_{12}$ & 0.88 & 48$^\circ$C & 2.8 & 1.51 & 0.8 (20$^\circ$C) \\
Anisole & C$_7$H$_8$O & 0.99 & 43$^\circ$C & 3.5 & 1.52 & 1.4 (15$^\circ$C)\\
Methylnaphtalene & C$_{11}$H$_{10}$ & 1.02 & 82$^\circ$C & 2 & 1.62 & 3 (20$^\circ$C)\\
LAB & -- & 0.86 & $\sim140^\circ$C & 0.013 & 1.48 & 5 (20$^\circ$C)\\
DIN & C$_{16}$H$_{20}$ & 0.96 & $\sim140^\circ$C & 0.005 & 1.56 &  6 (40$^\circ$C)\\ 
ortho-PXE & C$_{16}$H$_{18}$ & 0.99 & 167$^\circ$C & 0.0013 & 1.56 &  6 (40$^\circ$C)\\
Polysiloxane (TPTMTS) & -- & 1.07 & 230$^\circ$C & 5$\cdot10^{-7}$ & 1.58 & 54 (25$^\circ$C)\\
\hline
\end{tabular}
\end{center}
\end{table}

\subsubsection*{Linear alkyl benzene}
In neutrino physics, linear alkyl benzene (LAB) was first proposed in the context of the SNO+ experiment~\cite{Chen:2007ak}. Among its main advantages are high transparency, good compatibility with typical materials used for the detector vessels as acrylic, availability on large scales and low cost. In terms of safety, it has a flash point well above 100$^\circ$ and it is usually not defined as a hazardous material under transport regulations. For those reasons LAB was the scintillator basis of many experiments in the last 15 years. It was used on the multi ton scale in experiments as Double Chooz (veto scintillator)~\cite{DC}, Daya Bay~\cite{DB} or RENO~\cite{RENO}. It is also the solvent of choice for the upcoming JUNO experiment~\cite{JUNO}, which plans to use 20~ktons of LAB based scintillator. The absorption maximum around 260\,nm and emission maximum slightly below 300\,nm are similar to the ones of classical solvents such as xylene or PC. However, depending on the purity contaminations of bi-phenyls might effect the absorption and emission spectra above 350\,nm.   

\subsection*{Diisopropylnaphtalene}
A scintillator solvent with similar flash point as LAB is diisopropylnaphtalene (DIN). In liquid form it is typically available as a mixture of isomers. There are also specific isomers such as 2,6-DIN which are solid at room temperature. Compared to LAB it has a significantly higher density close to 1. This can be advantageous for mechanical stability in the case of nested detector vessels with a water-based neighboring volume, since in this case buoyancy forces are much weaker as compared to other solvents. Another advantage of DIN is that it offers one of the best performances in terms of pulse shape discrimination (PSD), similar to xylene based liquids. DIN was applied as admixture in the scintillator solvent in NEOS~\cite{NEOS:2016wee}, Stereo~\cite{Stereo} or Nucifer~\cite{NUCIFER:2015hdd} to optimize light yield and PSD capabilities. Commercial scintillator cocktails as the Ultima Gold series from Perkin Elmer also profit from the favorable DIN properties.

One of the disadvantages of DIN is the rather low transparency. The best literature value we could find for the attenuation length at a benchmark wavelength of 430\,nm is 4.2\,m after column purification~\cite{Song:2013vxa}. However, depending on the optical purity of the liquid before the final purification steps, the attenuation length can be in the 1\,m range only~\cite{StereoLS}. Moreover, DIN is classified as environmental hazard, which complicates transport and handling of the substance. Concerning the emission spectrum it is shifted towards higher wavelengths as compared to classical solvents with a peak around 340\,nm. Therefore the overlap to the absorption peaks of typical fluors as p-terphenyl or PPO is not ideal. 

\subsection*{Phenyl xylyl ethane}
In terms of density and safety characteristics phenyl xylyl ethane (PXE) is comparable to DIN. The molecule exists in the para-, meta- or ortho-form. Transparent and high purity batches were found in particular for the ortho-PXE isomer. The molecule is characterized by a high light yield, but has limited PSD capabilities. The transparency is typically better than DIN, but worse than for LAB. On the multi ton scale PXE was first proposed as back-up solvent for the Borexino experiment. In the Borexino counting test facility (CTF) it could be demonstrated that for thorium and uranium purity levels below 10$^{-16}$\,g/g can be achieved after column purification~\cite{PXEpaper}. Moreover, PXE was the scintillator basis in the Double Chooz experiment. To improve compatibility with the acrylic vessel it was mixed with about 80\% n-dodecane~\cite{DCLS}. Ortho-PXE was also contained in the scintillators of Stereo~\cite{Stereo} and Nucifer~\cite{NUCIFER:2015hdd}. Absorption and emission maxima are similar to those of LAB or classical solvents such as PC.

\subsection*{Methylnaphtalene}
One of the first proposals in the direction of a scintillator solvent with high flash point and density close to 1 was the use of 1-methylnaphtalene (MN)~\cite{1MN}. The vapor pressure is significantly higher than for the previously described chemicals and the flash point in between classical solvents and modern safe scintillators. One of the striking features of MN is the rather high light yield. The authors in~\cite{1MN} reported a scintillation pulse height 10\% above the one of a xylene based LS. The challenges of the MN approach are already pointed out in the same reference as well. Commercially available MN has often a yellowish color as received and needs to be purified before it is used in a LS. Column purification with acidic alumina oxide significantly improves the transmission, but careful handling of the sample is needed to maintain transparency on the scale of months or longer. To keep stability the liquid should be stored in the dark and air exposure has to be avoided~\cite{1MN}. In principle this is true for other organic solvents as well, however MN is more sensitive to reactions that cause yellowing than the safe solvents described above. It is probably for those reasons that MN was hardly used and implemented in large scale LS detectors. As for other naphtalene based molecules such as DIN, the emission spectrum is shifted towards longer wavelengths as compared to aromatics with one phenyl ring.    

\subsection*{Silicon oils}
Recently a new class of safe liquids was proposed for LS detectors with many interesting features, namely liquid polysiloxanes~\cite{Polysilo}. In terms of safety they even excel the properties of the aforementioned organic solvents. The flash point is typically well above 200$^\circ$C and the vapor pressure is extremely low. Therefore, these odorless liquids do not produce harmful vapors at all. Moreover, these compounds are chemically inert, which supports stability and good material compatibility. Basis for the chemical and thermal robustness of this material are the strong Si-O bonds in the main chain of the molecules. Formation of free radicals, which are a known source of yellowing processes in organic materials, is also reduced as compared to solvents with the weaker C-C bonds. There are no speciﬁc risks for health or environment as well. The viscosity of the polysiloxanes is rather high. This complicates filtering processes and has to be taken into account in the design of the liquid handling systems. But there is also the positive aspect that high viscosity liquids are less subjected to spills, another beneficial argument in terms of safety. There are silicon oils with viscosities of 10000\,mm$^2$/s at room temperature. Polysiloxane based scintillators were also proposed in solid forms~\cite{polysolid}.   

In this work we mainly concentrated on a specific polysiloxane liquid having the most promising characteristics based on previous studies~\cite{Polysilo}. The compound tetraphenyl-tetramethyl trisiloxane (TPTMTS) has a high molar fraction of phenyl groups and favorable absorption and emission characteristics providing high light yields. Moreover, it has a rather low viscosity as compared to other liquid polysiloxanes simplifying the liquid handling. It was difficult to get this material for a reasonable price at the scale of several liters from the typical companies selling laboratory chemicals. However, we found that some diffusion pump oils are mainly based on TPTMTS (DOWSIL$^{TM}$ 704, Wacker AN140 or Sindlhauser S-04V) which can be obtained at lower cost and with a shorter delivery time. However, the chemical purity could not be specified for the samples we received from Sindlhauser Materials. Moreover two other polysiloxanes were tested, 1,1,3,5,5-pentaphenyl-1,3,5-trimethyltrisiloxane (PPTMTS) and polymethylphenylsiloxane (PMPS). The PPTMTS was purchased as S-05V, as well from Sindlhauser Materials, and the PMPS from abcr GmbH.  

\subsection*{Solvent mixtures}
Another approach to improve the safety characteristics and material compatibility of LS is to mix them with more inert liquids as mineral oil or n-alkanes. In KamLAND~\cite{Kamland} and Double Chooz~\cite{DCLS} for example a mixture of about 80\% n-dodecane and 20\% aromatic fraction were used in the target liquids to improve the compatibility with the acrylic vessels. Recently there is also increasing interest in water based LS with an organic fraction of only several percent~\cite{Yeh:2011zz}.  

A novel technology in the field is the use of opaque scintillators~\cite{Cabrera:2019kxi}. A promising approach in this direction is the use of a wax based scintillator which is liquid around 40$^\circ$C and solidifies below 30$^\circ$~\cite{nowash}. With this material the risk of liquid spills and associated dangers can be completely avoided, but one can still profit from the advantages of a liquid at higher temperatures.   

\subsection{Primary wavelength shifters}
The energy transfer between the excited solvent molecules and the fluor molecules as acceptor is ideally radiationless. For an efficient energy transfer mechanism a good overlap between solvent emission and fluor absorption spectra is needed. Considering the emission peaks of the safe solvents introduced above the main fluor absorption should be in the $300-350$\,nm regime. Another important fluor property to achieve high light yields in the scintillator is the quantum yield which is typically around 90\%~\cite{Buck:2015jxa}. In order to reduce the self-absorption of the fluor emission a large Stokes shift, the difference between absorption and emission peak, can be useful.  

We studied the performance of several fluors in combination with the safe solvents. In particular the main focus was on the following candidates:

\begin{itemize}
\item PPO (2,5-Diphenyloxazol): This fluor is used in most of the currently running LS detectors, since it is highly transparent above 420~nm, radiopure and available at large amounts for reasonable cost. Moreover, the solubility in most solvents is very high. 
\item butyl-PBD (2-(4-biphenyl)-5-(4-tert-butyl-phenyl)-1,3,4-oxadiazole): This slightly larger molecule has a high quantum yield and is therefore expected to provide high light yields as well. 
\item BPO (2-(4-biphenyl)-5-phenyloxazole): BPO is known to provide high light yields and its absorption and emission spectra are shifted towards longer wavelengths as compared to PPO or butyl-PBD by about 20~nm. Therefore, it is expected to perform particularly well in a solvent like DIN due to the better overlay between spectra. However, a disadvantage of BPO is that it is more hazardous than the other fluors reported here.
\item PMP (1-Phenyl-3-mesityl-2-pyrazolin): The fluor PMP is known for its high Stokes shift reducing self-absorption. Therefore it might allow to be used in a binary system of just two components, solvent and fluor. In most other cases a secondary wavelength shifter is needed in large detectors.
\item NPO (2-(1-Naphthyl)-5-phenyloxazole): Another promising candidate is NPO with an absorption maximum of  334\,nm and an emission peak at 398\,nm~\cite{NPO}. 
\end{itemize}

\subsection{Secondary wavelength shifters}
The number of typical candidates used as secondary wavelength shifter (WLS) is much smaller than the one for the primary fluors. The most commonly used molecules here are bis-MSB (1,4-bis(2-methylstyryl)benzene) and POPOP (1,4-bis(5-phenyloxazol-2-yl)benzene). The properties and performance of the two is rather similar and we could not identify big advantages of one over the other concerning the optical properties. 

The most important features of the secondary WLS, which is typically added at small concentrations of about 10~ppm are a high quantum yield and favorable absorption and emission spectra. The absorption peak should overlap with the emission of the primary fluor around $350-400$\,nm and the light emission should match the wavelengths of highest sensitivity for the photomultiplier tube (PMT) used for light detection. This is commonly in the range of $400-450$\,nm. 

\section{Experimental setup}
\label{sec:experimental}
In the studies on safe LS reported in this article we investigated the transparency, light yield and PSD capabilities of the mixtures. For the transparency measurements a standard UV-Vis photospectrometer with 1~cm and 10~cm long cells was used. Light yield (LY) and PSD were measured in the same setup consisting of a scintillator cell with an attached PMT, which was read out by a dedicated DAQ system.  

\subsection{Scintillator module}
The scintillator cell was developed and first used in the context of an ionization quenching measurement in germanium (Ge) crystals~\cite{Bonhomme:2022lcz}. In this measurement, eleven LS modules were positioned around a Ge detector for a coincidence measurement of scattered neutrons inside the crystals. One of these modules is now used for LS characterization at MPIK and described below.   

\begin{figure}[htbp]
\centering 
\includegraphics[width=0.8\textwidth]{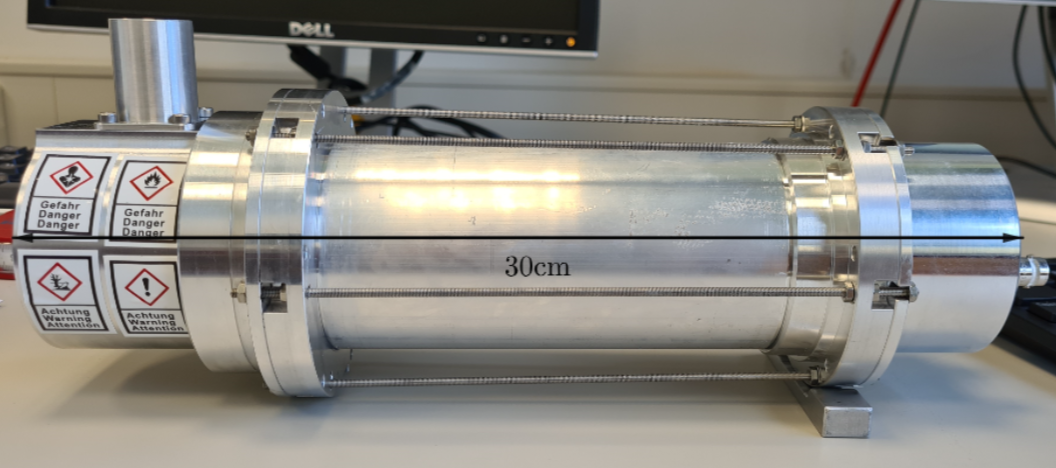}
\caption{\label{fig:1} The picture shows the scintillator module used to measure the relative light yield and pulse shape discrimination capabilities of the liquids.}
\end{figure}

The liquid is contained in a cylindrical PTFE container with a glass window. The liquid capacity of the small vessel is about 110~ml. PTFE was chosen due to its high reflectivity and excellent compatibility with most organic liquids. This LS cell is viewed by a 2 inch PMT, which is attached to the glass window of the cell using optical grease. The PMT in the module is an ETL9954B with a maximum quantum efficiency of about 27\% at wavelengths around 400~nm. PMT and LS cell are enclosed in a light tight aluminium housing (see Figure~\ref{fig:1}). At one end of this housing are the two connectors for the positive 1.5~kV high voltage and the signal cable. The signals from the PMT are acquired with a V1725 module from CAEN, allowing to record the signal waveforms at a sampling rate of 250 MHz and to compute the corresponding charge by integration. The module is controlled via the CoMPASS software.

The scintillator samples were filled into the cells via a little opening at the top of the PTFE body. Then the sample is flushed with argon or nitrogen gas for about 10 minutes to remove oxygen dissolved in the LS. No significant difference on the LY was observed for the use of the two gases.
For the characterization of the light signal, radioactive sources are positioned in front of the LS module. For the LY determination a $^{137}$Cs source was used which produces mono energetic gammas with an energy of 662~keV. The activity of the $^{137}$Cs source at the time of the measurements was about 130~kBq. To determine the PSD properties an AmBe source producing neutrons and gammas was utilized. In the alpha decays of $^{241}$Am, gammas with energies up to 1.015~MeV are emitted. The interaction of the alpha particles with the $^{9}$Be isotope produces excited states of $^{13}$C, which decay under emission of one neutron and in some cases a 4.4~MeV gamma. The neutron energies from the AmBe source extend up to 11~MeV, however compared to gamma or electron sources the scintillation signal is quenched by a factor of 2 to 3. The AmBe source used in the measurements produced about $7\cdot10^4$ neutrons per second.  

\subsection{Light yield determination}
\begin{figure}[htbp]
\centering 
\includegraphics[width=0.6\textwidth]{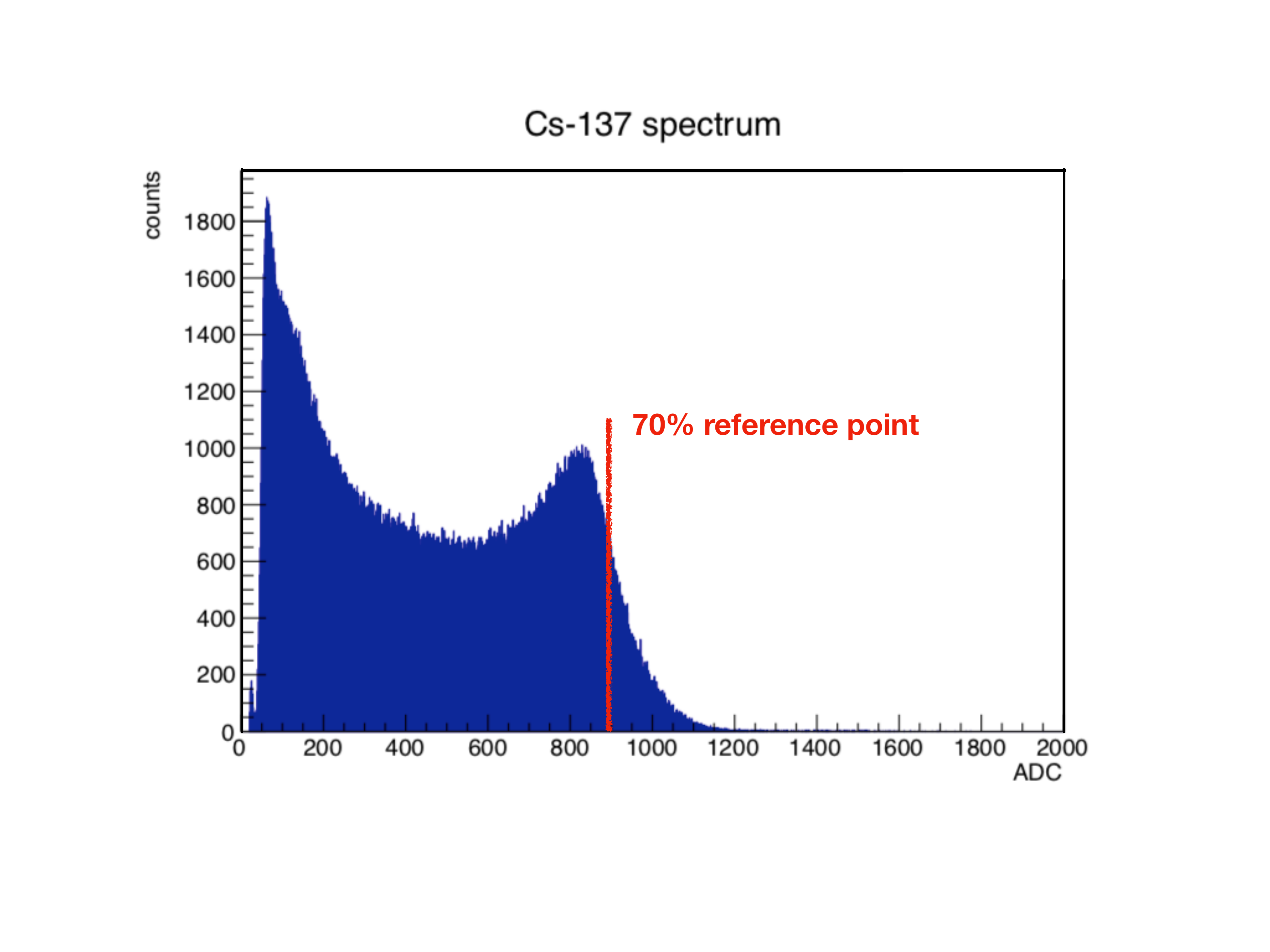}
\caption{\label{fig:2} A typical Compton spectrum as measured with the $^{137}$Cs source is shown with the ADC channel number on the x-axis. The reference point at 70\% from the peak maximum is indicated in red.}
\end{figure}

To determine the LY of the samples a relative comparison of the scintillator response was performed at a defined reference point of the Cs energy distribution. Because of the small volume of the cells, the Compton edge is used. As an example the Compton spectrum for one of our scintillator standards is shown in Figure~\ref{fig:2}. We tested two methods for the LY determination. The first one is taking the ratio of the ADC channels determined around the Compton edge of the spectrum measured with the $^{137}$Cs source. The difficulty here is that due to the finite energy resolution the Compton edge is smeared. This smearing depends on the LY of the samples. There were several attempts in the literature to identify the best position in the Compton spectrum for a relative comparison between samples. This position is typically defined as the ADC unit for which the count rate in the high energy tail of the spectrum drops to a certain percentage from the maximum of the Compton peak. The suggested values in the literature for this optimum position ranges from $66-89$\%~\cite{Safari:2016ypc}, however this optimum depends on the detector design and the LS characteristics. To estimate the associated systematics we extracted the relative LY of different scintillator samples as compared to a LS standard (Eljen EJ301) for 3 different reference points at 70\%, 80\% and 90\% from the maximum. The differences in the extracted relative light yield between reference points were all below 2\%.

The second way to determine the relative LY is to make use of the Compton backscattering method. Here, the gamma source is positioned in between the LS sample and a second LS detector. An event is only recorded in case there is a coincidence signal in the second detector. Therefore, only gammas with a backscattering angle around 180 degrees corresponding to the maximum recoil energy at the Compton edge are included. This procedure allows to determine a well defined reference point which can be obtained by a symmetric gaussian fit. However, the statistics in this method is reduced. If the cells are positioned too close to each other a rather wide range of scattering angles is allowed, adding additional systematic uncertainties. On the other hand, if the distance between source and cells is increased the statistics is highly deteriorated. Due to those limitations we adopted the simpler Compton method for the measurements reported in this article. For one sample we compared the LY obtained for the two methods and found very similar values for the case of the 70\% reference point in the Compton method. Therefore we used this 70\% as reference point (see Figure~\ref{fig:2}).

\subsection{Pulse shape discrimination}
Scintillation events induced by photons (Compton electrons) and neutrons (proton recoils) can be separated by the different decay times. The different ratios of charge integrals of a short gate window covering the tail of the pulse ($Q_{tail}$) and a long gate period over the full signal ($Q_{tot}$), allow to distinguish between events with different ionization density. Therefore, we define the PSD parameter $r=Q_{tail}/Q_{tot}$.

\begin{figure}[htbp]
\centering 
\includegraphics[width=0.8\textwidth]{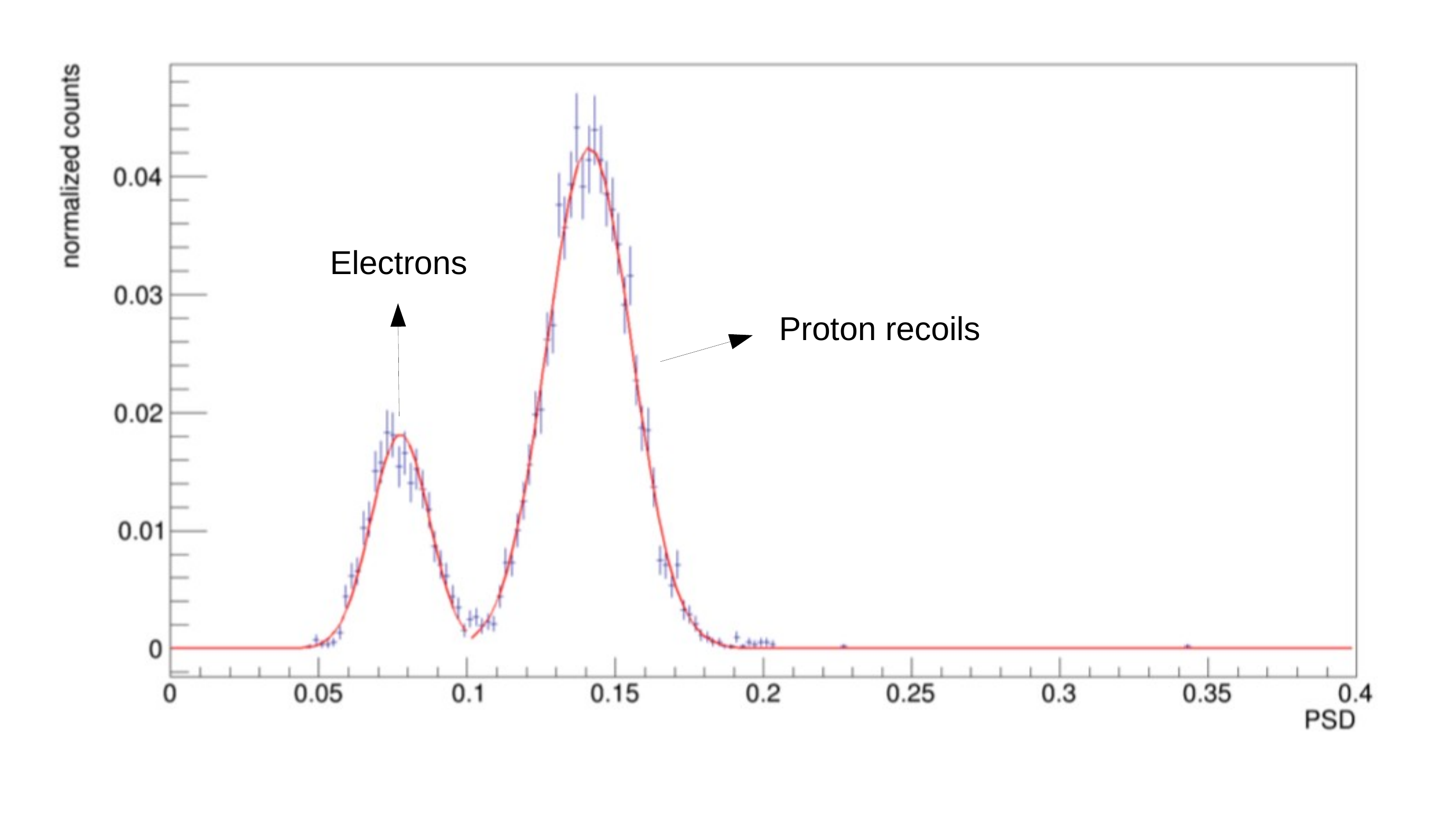}
\caption{\label{fig:3} Distribution of the PSD parameter for a LS standard (LAB, PPO and POPOP) with a FoM$_{max}$ of 1.18.}
\end{figure}

To quantify the separation power between the nuclear and electronic recoil population for a given energy interval the figure of merit (FoM) can be defined as:

\begin{equation}
FoM := \frac{1}{2\sqrt{2ln(2)}}\frac{|\mu_n - \mu_e|}{\sigma_n+\sigma_e}
\end{equation}

where $\mu_n$ and $\sigma_n$ are the central value and the standard deviation of the nuclear recoil, $\mu_e$ and $\sigma_e$ of the electronic recoil population accordingly. An example distribution is shown in Figure~\ref{fig:3}. The value for the FoM is energy dependent, so the asymptotic maximum (FoM$_{max}$) obtained at high energy is used to compare different scintillators.

\subsection{Systematic uncertainties}
A combined set of systematic uncertainties can be estimated by studies of the reproducibility of the results. By measuring a sample several times on different days, each time starting with a empty scintillator cell, the variance of the results indicates the contributions of several systematic effects. These are for example the efficiency of the oxygen removal, filling level in the cell, temperature effects, HV stability, sample preparation effects etc. Method related systematics such as the reference point in the Compton spectrum, variations in source position, fluctuations in the light collection due to differences of the emission spectra are estimated to be small compared to the systematics affecting the reproducibility. From these studies the relative systematic uncertainty on the LY determination was estimated to be 2.4\%. Based on the reproducibility of several different measurements of the same sample, the uncertainty of the FoM of the PSD was estimated to be only 1\%. However, from a comparison with a completely different setup~\cite{StereoLS} we estimate an uncertainty from the method on the FoM$_{max}$ ratio for two samples which is about 5\%.  

\section{Results and discussion}
\label{sec:results}

\subsection{Liquid scintillator standards}
Commercially available LS are typically used as standards to perform relative LY and PSD measurements, since an absolute determination would require a detailed simulation and understanding of the detector setup. Nevertheless, a comparison of measurements between different setups even under the use of similar standards can be difficult, since the methods and associated systematic uncertainties or biases might vary significantly between laboratories. Therefore a comprehensive comparison of different LS measured in an identical setup is beneficial.

Popular LS standards with widely reported LY and excellent PSD properties are EJ301 (Eljen Technology), BC501A (Bicron) and NE213 (Nuclear Enterprises). These LS are all chemically very similar and have basically identical properties. They are based on the classical solvent xylene with an admixture of naphthalene and therefore not suitable for applications with strong safety requirements. The LY of these standards as stated by the suppliers is 78\% of anthracene, which itself produces about 17400 photons/MeV. Given those numbers the LY of EJ301 and BC501A would be approximately 13500 photons/MeV. To avoid aging effects it is very important to control the storage conditions of the LS standards. Exposure to oxygen and light have to be avoided as much as possible. On the other hand extensive nitrogen purging might lead to significant evaporation of the volatile xylene. In this case the xylene to naphtalene ratio would decrease, which can lead to a change of the optical properties as well. For the measurements reported in this article, we also did a comparison to the performance of a p-xylene LS without naphtalene mixed in our laboratory. The LY of this liquid was the same as for the commercial products such as EJ301, but without the naphtalene the PSD performance is lower.

Another LS used in many laboratories is Ultima Gold F (UG-F) from Perkin Elmer. This is a DIN based LS and therefore advertised as safe liquid. The LY measured for our UG-F sample was $84\pm2$\% compared to EJ301. The PSD performance of UG-F was slightly worse than the one for EJ301 as well. 

\subsection{Safe scintillators}
Nowadays, the most commonly used safe LS in large neutrino detectors is a mixture of LAB as solvent, PPO as fluor and a secondary WLS (typically bis-MSB or POPOP). The optical and PSD properties of LAB based LS were reported in many publications~\cite{Lombardi:2013nla, LABLY, Wurm:2010ad, JUNO:2020bcl, SNO:2020fhu}. As for other solvents the LY starts to saturate at PPO concentrations of a few g/l~\cite{Ye:2015ska, CPL}. Typical concentrations of the secondary WLS are from $5-20$\,mg/l. Therefore we chose 4\,g/l PPO and 10\,mg/l POPOP as our standard fluor concentrations for our LAB samples. Compared to EJ301 we determined a LY for this LAB scintillator of $77\pm1.9$\% corresponding to 60\% of anthracene. This is more than the $\sim$50\% reported for LAB LS in~\cite{LABLY}, but $10-15$\% less than the number estimated in~\cite{SNO:2020fhu}. One of the drawbacks of LAB is the limited PSD performance. A comparison of LY and PSD for xylene based scintillators to several selected safe LS is shown in Table~\ref{tab2}.

\begin{table}[h]
\caption[LY_PSD]{The light yields relative to anthracene are listed based on relative measurement comparing to EJ301. The value for the commercially available standards in the first line is taken from the data sheets. To calculate the number of photons/MeV in column 3 a value of 17400 photons/MeV is assumed for anthracene. The last column is the ratio of FoM$_{max}$ compared to the LAB standard. The estimated uncertainty of this ratio is 5\%. All samples except the standard in the first line were mixed in our laboratory. If not stated otherwise we added 4\,g/l PPO and POPOP as secondary WLS. \label{tab2}}
\begin{center}
\begin{tabular}{lccc}
Scintillator & LY (\% anthracene) &  LY (photons/MeV) & PSD \\
\hline
EJ301/BC501A/NE213 & 78 & 13570 & 2.00 \\
p-Xylene & $77.2\pm1.9$ & 13430 & 1.45 \\
LAB & $60.2\pm1.4$ & 10470 & 1.00\\
DIN & $70.0\pm1.7$ & 12190 & 1.70\\
ortho-PXE & $64.8\pm1.6$ & 11270 & 1.03\\
1-MN (no PPO, 0.5\,g/l POPOP) & $73.4\pm1.8$ & 12770 & 1.31 \\
TPTMTS (1.2\% butyl-PBD, POPOP) & $58.4\pm1.4$ & 10170 & 0.92\\
\hline
\end{tabular}
\end{center}
\end{table}

Higher LY and better PSD properties can be achieved with DIN (see Table~\ref{tab2}). However, even after purification we did not succeed to obtain attenuation lengths of more than 2\,m at a reference wavelength of 430\,nm, although better values of $3.5-4.2$\,m were reported in~\cite{Song:2013vxa}. Also the DIN based UG-F sample from Perkin Elmer had an attenuation length below 1\,m. The stronger absorbance as compared to LAB is obviously a highly relevant limitation for large scale detectors in which the scintillation light has to propagate several meters in the liquid before it hits a light sensor. To profit from the high LY and strong PSD performance of DIN and still keeping reasonable transmission it is sometimes used as admixture in LAB based LS~\cite{StereoLS}. We found that the LY dependence on the DIN/LAB mass ratio is consistent with a linear dependence. Even with small admixtures of DIN to LAB of 5-10\,wt.\% significant improvements on the PSD behavior could be achieved consistent with the results reported in~\cite{Kim:2015pba} 

For the o-PXE sample we measured a LY higher than LAB, but lower than DIN and found a FoM for the PSD which is similar to the one of LAB. In terms of transparency after purification PXE is in between DIN and LAB. To our knowledge the highest reported attenuation length at 430\,nm was 12\,m~\cite{PXEpaper}. For the samples we received in the last years typical attenuation lengths were about 2-3\,m before and 6-8\,m after Al$_2$O$_3$ column purification.    

For the solvent MN the primary fluor is less important, since the emission spectrum is shifted towards longer wavelengths. We found good performance in a binary system of just MN as solvent and POPOP as the sole WLS. Such a two component system can have some advantages. First, it simplifies production and second it could help the radiopurity, since the concentration of radioactive impurities is typical higher in the fluors than in the solvents. The attenuation length of 1-MN as received was just 0.2\,m and could be improved to 1.1\,m after Al$_2$O$_3$ column purification. With the prurified MN and a POPOP concentration of 0.5\,g/l a LY of 95\% compared to the xylene LS could be achieved which is the highest value we found among all the safe solvents. Even at only 0.2\,g/l POPOP the LY was still 88\% compared to the xylene LS. The addition of PPO to the MN/POPOP system did not improve the LY. The PSD properties for MN are better than for LAB or PXE, but worse than for DIN as shown in Table~\ref{tab2}.  

The measurements on the polysiloxanes concentrated on TPTMTS (Sindlhauser S-04V). It has a lower viscosity and higher LY than most other polysiloxanes~\cite{Polysilo}. The high LY can be explained by the favorable emission spectrum and a high fraction of phenyl groups in the liquid. The fluor concentrations needed for polysiloxanes are higher than for the other solvents discussed in this article. This could be related to the higher viscosities which reduce energy transfer from processes in which diffusion plays a role. With 1.2\% butyl-PBD and POPOP as WLS a LY comparable to our LAB standard could be achieved. The authors in~\cite{Polysilo} even reported a LY for a TPTMTS scintillator which is close to the xylene based EJ309 (Eljen Technology) corresponding to 80\% anthracene. The lower LY measured in our case might be related to the lower purity of the S-04V. Nevertheless, the attenuation length measured for an unpurified S-04V sample was more than 5\,m. The PSD properties of the S-04V scintillator was similar to the one of LAB. With NPO or BPO instead of butyl-PBD as fluor 95\% of the LY could be achieved at half of the fluor concentration (0.6~wt.\%) at similar PSD performance. 

We also tested higher viscosity polysiloxanes, namely PPTMTS (S-05V) and PMPS. The high viscosity, 175\,mm$^2$/s for PPTMTS and 500\,mm$^2$/s for PMPS  complicates liquid handling or filtering. However, such high viscosity media have the advantage that they are much less sensitive to spills in case of small detector leaks. Moreover, high metal-loading might be achieved in such materials using inorganic micro- or nanocrystals suspended in the gel phase. The LY of PPTMTS (PMPS) was measured to be 9\% (27\%) lower as compared to TPTMTS consisted with the results in~\cite{Polysilo}. 

\section{Discussion}
\begin{figure}[htbp]
\centering 
\includegraphics[width=0.6\textwidth]{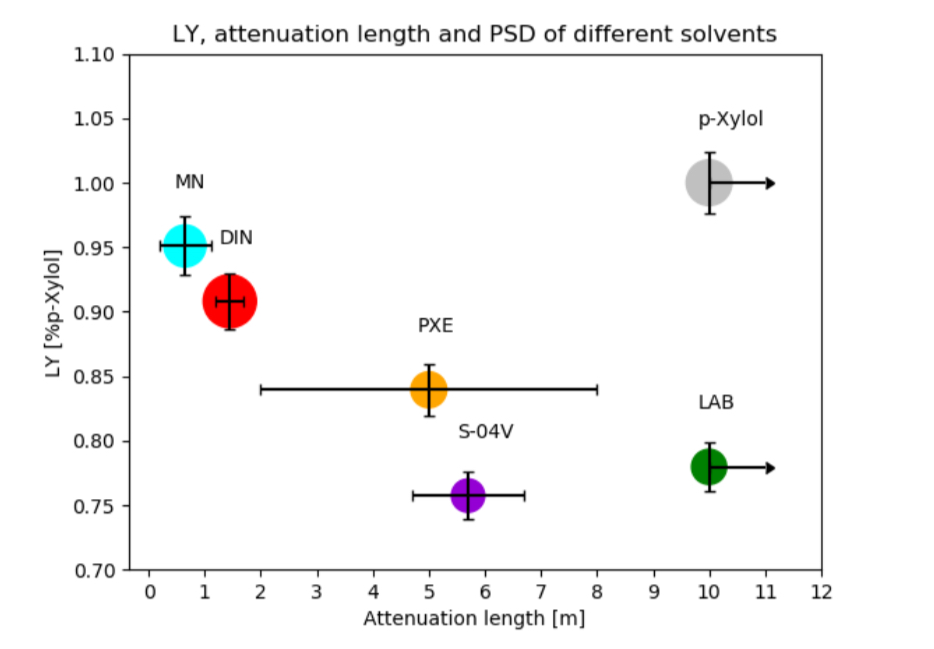}
\caption{\label{fig:4} The light yield of several safe scintillators is plotted versus the attenuation length and compared to a xylene based standard. The size of the dots indicates the PSD performance, the larger the diameter the better the PSD capabilities.}
\end{figure}

In many experiments using liquid scintillators, safety aspects related to the organic solvent are highly relevant. In particular for ton scale volumes and at sites with special safety requirements there is a need for high flash point and non-toxic solvents such as LAB, DIN, PXE or polysiloxanes. In this article optical and PSD properties of different safe liquid scintillators were compared in the same setup. Some of the dominant systematic uncertainties related to the methods cancel in these relative measurements. A summary of the obtained results is shown in Figure~\ref{fig:4}. This plot might be used as a guideline in the design phase of particle detectors to identify the most appropriate scintillator composition. 

The preferred solvent choice in many running and upcoming neutrino experiments is LAB. As seen in Figure~\ref{fig:4} it is characterized by its favorable transparency. The light yield is good, but not outstanding, and the PSD capabilities are limited. It is available on large scales at moderate cost. Therefore LAB is suitable for very large detectors in which scintillation light has to propagate several meters and PSD performance is not among the most critical requirements. In smaller detectors asking for high light yield and strong PSD, DIN is probably the better solvent candidate. To find a compromise between transparency and PSD there is also the option to use mixtures of LAB and DIN. Another advantage of DIN is the higher density close to 1 similar as PXE. Therefore DIN or PXE could be used in vessels surrounded by water without creating strong buoyancy forces. PXE is in between LAB and DIN for light yield and transmission properties. The highest light yield and good PSD properties were found for the solvent methylnaphtalene. This solvent can be used in a binary system with just some POPOP as fluor. The drawback of methylnaphtalene is its short attenuation length and the poor stability. 

Finally the properties of liquid polysiloxanes were studied which offer a class of solvent candidates, so far not under the radar of most particle physics experiments. Those high viscosity liquids offer outstanding safety characteristics, since they are non-hazardous, have very low vapor pressures and flash points above 200$^\circ$C. Transmission of the most promising polysiloxane candidates before any purification step is better than for comparable DIN or PXE samples. Light yield and PSD are almost on the level of LAB. The density is the highest among all the investigated solvents. Therefore it is also a good candidate for an active veto system surrounding the detector, since it provides more shielding against external radioactivity. The higher average Z in the silicon oil reduces the mean free path length of low energy gammas due to the increased probability for the photoelectric effect. Polysiloxane based scintillators might be used in an environment with highest safety standards or in cases in which there is demand for gel-like scintillator materials.

\end{document}